\documentclass[pra,12pt,tightenlines]{revtex4}

\usepackage{bm}

\usepackage{graphicx}

\begin{document}

\def\ket#1{|#1\rangle}
\def\bra#1{\langle#1|}
\def\tr{{\rm tr}}
\def\p{{\bf p}}
\def\yes{{\rm yes}}

\newcommand{\vc}[2]{\left(\begin{array}{c}{\!\!#1\!\!}\\{\!\!#2\!\!}\end{array}\right)}

\title{Quantum theory from four of Hardy's axioms}

\author{R\"udiger Schack \\
\it Department of Mathematics, Royal Holloway, University of London \\ 
Egham, Surrey TW20~0EX, UK \\ 
E-mail: r.schack@rhul.ac.uk \\
$\;$ \\ $\;$}

\begin{abstract}
  In a recent paper [e-print quant-ph/0101012], Hardy has given a derivation
  of ``quantum theory from five reasonable axioms.'' Here we show that Hardy's
  first axiom, which identifies probability with limiting frequency in an
  ensemble, is not necessary for his derivation.  By reformulating Hardy's
  assumptions, and modifying a part of his proof, in terms of Bayesian
  probabilities, we show that his work can be easily reconciled with a Bayesian
  interpretation of quantum probability.
\end{abstract}

\pacs{}

\maketitle

$\;$ \section{Introduction}

In Bayesian probability theory \cite{Bernardo1994,Kyburg1980}, probabilities
are not objective states of nature, but rather are taken to be degrees of
belief that determine an agent's decisions in the face of uncertainty.
It can be shown that degrees of belief must obey the usual rules of the
probability calculus if the agent's decisions are rational (for
references and a summary of the argument, see \cite{Caves2002a}). In a Bayesian
framework, probabilities and measured frequencies are strictly separate
concepts. This leads to conceptual clarity in statements that involve both
probabilities and frequencies. Furthermore, adopting the Bayesian viewpoint
has important practical consequences in the field of statistics
\cite{Bernardo1994,Malakoff1999}.

If the Bayesian interpretation is applied to quantum mechanical probabilities,
one is led naturally to the viewpoint that quantum states represent states of
belief. This viewpoint is attractive for many reasons. For instance, it
eliminates the difficulties associated with regarding quantum state collapse
as a real physical process. Within the Bayesian framework, one can account
effortlessly for the tight connection between measured frequencies and the
probabilities obtained from the quantum probability rule \cite{Caves2002c}.
The Bayesian approach has led to new mathematical results
\cite{Caves2002b,Caves2002a}, a better understanding of prior information in
quantum tomography \cite{Schack2001a}, and an optimized entanglement
purification protocol \cite{Brun2001a}.

Hardy \cite{Hardy-0101} (see also \cite{Hardy-0111}) has recently given a
derivation of the mathematical structure of quantum theory from five simple
axioms. In his first axiom, Hardy identifies probability with measured
frequency in the limit of an infinite number of repetitions of a given
experiment. In Hardy's formulation, a quantum state is a property of a
preparation device. This is a problematical notion. Attempts to base
probability theory on a definition of probability as frequency in infinite
ensembles \cite{VonMises1957} have largely failed (see, e.g.,
\cite{vanFraassen1977,Jaynes2003}). For instance,
without further complicating assumptions, a relative frequency specified for
an infinite ensemble does not in any way restrict the corresponding frequency
for a finite subensemble. Furthermore, attaching the notion of a quantum state
to a preparation device appears to limit quantum theory to the description of
laboratory experiments. But surely one would want to assign a quantum state,
e.g., to a light pulse arriving from a distant star.

The details of Hardy's mathematical proof turn out to be mostly independent of
the specific assumptions of his first axiom. In the present paper, we show
that it is indeed possible to reformulate Hardy's derivation in such a way
that the axioms refer to Bayesian probabilities for the outcomes of
measurements performed on a single physical system (see also Hardy's remarks
at the end of section 6.1 of \cite{Hardy-0101}). In Sec.~\ref{sec:Hardy}, we
briefly review Hardy's basic setup and axioms. In Sec.~\ref{sec:Bayesian}, we
provide a Bayesian reformulation of the problem and explain how Hardy's proof
can be modified accordingly. In Sec.~\ref{sec:discuss}, we conclude by showing
how the connection between probabilities and measured frequencies is
recovered in our formulation.

$\;$ \section{Hardy's setup}  \label{sec:Hardy}

In \cite{Hardy-0101},  Hardy considers the following situation. An
experimenter has a preparation device, a transformation device, and a
measurement device.  Associated with each preparation is a {\it state},
``defined to be (that thing represented by) any mathematical object that can
be used to determine the probability associated with the outcomes of any
measurement that may be performed on a system prepared by the given
preparation.''  If a physical system is incident on the measurement device, it
outputs a number $l$, where $l=1,\ldots,L$. If no physical system is incident
on the measurement device, it outputs the number 0.

Hardy then defines a {\it probability measurement\/} in the following way.  A
given measurement is performed on an ensemble of $n$ systems each prepared by
a given preparation device. Then the number of times, $n_+$, is recorded that a
particular outcome $l_1$, or subset of outcomes $S_1\subseteq\{1,\ldots,L\}$,
is observed.  The measured probability is then defined as
\begin{equation}
{\rm prob}_+ = \lim_{n\rightarrow\infty} {n_+\over n} \;.
\end{equation}

It is then assumed that there exists a minimum number, $K$, of appropriately
chosen probability measurements that completely specify the state of the
system. These $K$ probabilities can be represented by a column vector
\begin{equation}
 \p = (p_1,p_2,\ldots,p_K)^T \;,
\end{equation}
which  represents the state. The result of any probability
measurement can be inferred from the vector $\p$. The number $K$ is called
the {\it number of degrees of freedom\/} of the system. It follows from
the axioms below that the set of states is convex; pure states are defined
as the extremal points of this convex set.

Finally, the {\it dimension}, $N$, of the system is defined as the maximum
number of states that can be distinguished reliably in a single-shot
measurement. Using these terms, Hardy derives the usual Hilbert-space 
formulation of quantum theory from the following five axioms, quoted verbatim
from \cite{Hardy-0101}.
\begin{description}
\item[Axiom 1] {\it Probabilities}.  Relative frequencies (measured by
taking the proportion of times a particular outcome is observed)
tend to the same value (which we call the probability) for any case
where a given measurement is performed on a ensemble of $n$ systems
prepared by some given preparation in the limit as $n$ becomes infinite.
\item[Axiom 2] {\it Simplicity}. $K$ is determined by a function of
$N$ (i.e. $K=K(N)$) where $N=1,2,\dots$ and where, for each
given $N$, $K$ takes the minimum value consistent with the axioms.
\item[Axiom 3] {\it Subspaces}. A system whose state is constrained to
belong to an $M$
dimensional subspace (i.e. have support on only $M$ of a set of $N$ possible
distinguishable states) behaves like a system of dimension $M$.
\item[Axiom 4]  {\it Composite systems}. A composite system consisting of
subsystems $A$ and $B$ satisfies $N=N_AN_B$ and $K=K_AK_B$.
\item[Axiom 5] {\it Continuity}. There exists a continuous reversible
transformation on a system between any two pure states of that
system.
\end{description}

These axioms are stated in a manifestly frequentist language. Axiom 1 defines
probability in terms of limiting frequency, and the number of degrees of
freedom $K$, 
defined explicitly in terms of frequency measurements, has a central
position in both axioms 2 and 4. Nevertheless, a Bayesian formulation of
Hardy's program turns out to be straightforward.

$\;$ \section{The Bayesian setup}  \label{sec:Bayesian}

The Bayesian setup we are about to describe differs from Hardy's setup in the
following ways. In the Bayesian formulation, it will not be necessary to refer
to preparation devices or ensembles. Everything is expressed in terms of
single physical systems. The concept of a probability measurement is not
needed (see \cite{Caves2002b} for a Bayesian account
of what it means to effectively measure a quantum probability in a laboratory
experiment). Axiom 1 can be eliminated. 

Our primitives are physical systems, transformation devices and measurement
devices. As before, the non-null outcomes of a measurement device are labeled
$l=1,\ldots,L$. We define a special class of  measurements, so-called {\it
  yes-no measurements}, that have only two outcomes, which we label yes and
no.  E.g., for a given measurement device, the questions
``is the outcome equal to $l_1$?'' and ``is the outcome in the set
$S_1\subseteq\{1,\ldots,L\}$?'' define yes-no measurements. 

The state of a system is now defined to be {\it any mathematical object that
  summarizes a physicist's state of belief about a system in that it can be
  used to determine the probabilities associated with the outcomes of any
  measurement that may be performed on the system.} In this definition,
probability means the physicist's degree of belief about the outcome of a
measurement performed on a single system. Degrees of belief acquire an
operational definition in decision theory and can be shown to obey the usual
probability rules (see \cite{Caves2002a} for details and references).

We now assume that there exists a number of yes-no measurements such that the
probabilities for their outcomes determine the state fully. Let $K$ be the
minimum number of such yes-no measurements, and fix a set of $K$ such
measurements, the {\it fiducial measurements}. As before, the state is then
given by the probabilities assigned to the yes outcomes of the fiducial
measurements, i.e., by a vector $\p=(p_1,\ldots,p_K)^T$. For any yes-no
measurement there exists a function $f$ that maps any state $\p$ to the
probability for the yes outcome if the measurement is performed on a system
to which $\p$ is assigned. This can be expressed as
\begin{equation}
\Pr({\rm yes})=f(\p) \;.
\end{equation}
Finally, as before, the {\it dimension}, $N$, of the system is defined as the
maximum number of states that can be distinguished reliably in a single
measurement. 

It turns out that most parts of Hardy's proof are unaffected by our
reformulation \cite{Hardy-private}. Wherever
Hardy refers to a probability measurement, we refer instead to ``probability
assigned to the yes outcome of a yes-no measurement''. The only exception is
the part of the proof that uses axiom 1 explicitly, i.e., sections 6.4 and 6.5
of \cite{Hardy-0101}.

Section 6.4 of \cite{Hardy-0101} introduces the function $f$ defined above,
and derives the inequality $0\le f(\p)\le1$ from the assumption that
probabilities are measured frequencies. We get this inequality from the
assumption that the probabilities assigned by a physicist are degrees of
belief and therefore obey the laws of probability. Since $f(\p)=\Pr(\yes)$, it
follows that $0\le f(\p)\le1$. It is worth pointing out that the Bayesian
derivation of this inequality does not depend on the notion
of repeated trials and is therefore completely independent of the frequentist
derivation \cite{Caves2002a}.

Section 6.5 of \cite{Hardy-0101} introduces the idea of a mixture of two
quantum states, which is then used to derive linearity of the quantum
probability rule and quantum transformations. Hardy defines a mixture as an
ensemble consisting of a fraction $\lambda$ of systems prepared in a state
$\p_A$ and a fraction $1-\lambda$ of systems prepared in a state $\p_B$. This
construction cannot be used in our Bayesian approach, which refers only to a
single system, not a large ensemble of systems. In particular, in the
Bayesian approach, the mixing parameter $\lambda$ cannot be interpreted as 
a limiting frequency of systems prepared in a particular way. A different proof
of linearity is therefore required.

Our alternative derivation of linearity is based on the idea of conditioning,
which is central to Bayesian theory in general.  Assume that $\p_A$ and $\p_B$
are possible states for a given system. Then we can imagine a situation in
which a physicist's state assignment depends on some event $E$.  The event $E$
could be the outcome of a previous measurement, or some other piece of
information that affects his state assignments. If he knew that $E$ was true,
he would make the state assignment $\p_A$, and if he knew that $\neg E$ was
true, he would make the state assignment $\p_B$. We now assume that he
does not know the truth value of $E$.  Instead, he assigns the probabilities
$\Pr(E)=\lambda$ and $\Pr(\neg E)=1-\lambda$ to the events $E$ and $\neg E$,
and makes the state assignment $\p_C$.

If we now apply the function $f$ for a given yes-no measurement to the state
$\p_A$, we obtain the conditional probability for the outcome yes, given that
$E$ is true,
\begin{equation}
f(\p_A) = \Pr(\yes|E) \;.
\label{eq:fpa}
\end{equation}
Applying $f$ to the state $\p_B$ gives the conditional probability for the
outcome yes, given that $\neg E$ is true, 
\begin{equation}
f(\p_B) = \Pr(\yes|\neg E) \;.
\label{eq:fpb}
\end{equation}
Finally, applying $f$ to the state $\p_C$ gives the unconditional probability for the
outcome yes,
\begin{equation}
f(\p_C) = \Pr(\yes) \;.
\label{eq:fpc}
\end{equation}
Since we have assumed that the physicist's probability assignments are 
Bayesian degrees of belief, they must obey the usual probability rules (see
above). In particular, they obey the law of total probability,
\begin{equation}
\Pr(\yes)  =  \Pr(\yes|E)\Pr(E)+\Pr(\yes|\neg E)\Pr(\neg E) \;.
\end{equation}
By substituting Eqs.~(\ref{eq:fpa}--\ref{eq:fpc}) and the definition of
$\lambda$, we obtain
\begin{equation}
f(\p_C) = \lambda f(\p_A) + (1-\lambda) f(\p_B) \;.
\label{eq:totalprob}
\end{equation}
This is the same equation that Hardy derives in section 6.5 of
\cite{Hardy-0101}. Following Hardy, we can now apply Eq.~(\ref{eq:totalprob})
to the $K$ fiducial measurements. For the $k$-th fiducial measurement, $f(\p)$
is the $k$-th component of $\p$. Writing the $K$ resulting equations in vector
form, we obtain 
\begin{equation}
\p_C = \lambda \p_A + (1-\lambda) \p_B \;,
\end{equation}
which can be combined with Eq.~(\ref{eq:totalprob}) to give 
\begin{equation}
f(\lambda\p_A+(1-\lambda)\p_B) = \lambda f(\p_A) + (1-\lambda) f(\p_B) \;.
\label{eq:convexlin}
\end{equation}
This establishes convex linearity of the function $f$.  For a different
Bayesian derivation of Eq.~(\ref{eq:convexlin}), see \cite{Fuchs2002a}, which
builds on the theory of quantum Bayesian updating introduced in
\cite{Fuchs-0205}.

$\;$ \section{Discussion}  \label{sec:discuss}

In the previous section, we have seen that most of Hardy's derivation of
quantum theory remains valid if probabilities are given a Bayesian
interpretation. In the Bayesian formulation, Hardy's frequentist axiom 1 can
be omitted. Linearity now follows from the basic setup, where quantum states
are defined as compendia of probabilities for the outcomes of arbitrary
single-shot yes-no measurements. In this sense, quantum theory can be derived
from the last four of Hardy's five axioms.

It may seem, however, that something important is lost in the Bayesian
approach. Hardy's version of quantum theory makes statements about actual
frequencies measured in a laboratory, which are conspicuously absent from the
Bayesian formulation given above. We will now review an almost trivial
argument that establishes a tight connection between Bayesian quantum state
assignments and measured frequencies.

Suppose an experiment consisting of the preparation of a system and a
subsequent yes-no measurement is repeated $n$ times. Assume that the
experimenter assigns the $n$-fold tensor product state
\begin{equation}
\rho^{\otimes n} \equiv \rho\otimes\rho\otimes \cdots \otimes \rho
\end{equation}
to the $n$ copies of the system, where $\rho$ is a single-system density
operator. Suppose the single-system measurement is described by the projection
operators $P_{\yes}$ and $P_{{\rm no}}=1-P_{\yes}$. The probability for yes in
the first measurement is then $q\equiv\Pr(\yes)=\tr(\rho P_{\yes})$. The
probability for $k$ yes outcomes and $n-k$ no outcomes in $n$ repetitions of
the experiment is easily found to be
\begin{equation}
\Pr(k) = {\vc n k} q^k (1-q)^{n-k} \;,
\end{equation}
which for large $n$ is strongly peaked near $k/n=q$. The probability that the
measured frequency is near $q$ approaches 1 as $n$ tends to infinity.  The
Bayesian starting point of regarding probability and measured frequency as two
separate concepts thus leads to a transparent and tight connection
between quantum states and measured frequencies. Nothing is lost by abandoning
the {\it a priori\/} identification of probabilities with measured
frequencies.

$\;$ \section*{Acknowledgments}

I would like to thank Chris Fuchs for stimulating discussions and 
important suggestions. 


\begin{thebibliography}{10}

\bibitem{Kyburg1980}
{\em Studies in Subjective Probability}, 2nd edition, 
edited by H.~E. Kyburg and H.~E. Smokler 
(Robert~E. Krieger Publishing, Huntington, NY, 1980).

\bibitem{Bernardo1994}
J.~M. Bernardo and A.~F.~M. Smith, {\em Bayesian Theory\/} (Wiley, Chichester,
  1994).

\bibitem{Caves2002a}
C.~M. Caves, C.~A. Fuchs, and R. Schack, 
``Conditions for compatibility of quantum state assignments'',
to appear in {\it Phys.\ Rev.\ A\/} (2002), e-print quant-ph/0206110.

\bibitem{Malakoff1999}
D. Malakoff, ``Bayes offers a `new' way to make sense of numbers'',
{\it Science\/} {\bf 286},  1460  (1999).

\bibitem{Caves2002c}
C.~M. Caves, C.~A. Fuchs, and R. Schack, 
``Quantum probabilities as Bayesian probabilities'',
{\it Phys.\ Rev.\ A\/} {\bf 65},
  art.~no.~022305 (2002).

\bibitem{Caves2002b}
C.~M. Caves, C.~A. Fuchs, and R. Schack, 
``Unknown quantum states: The quantum de Finetti representation'',
{\it J. Math.\ Phys.} {\bf 43},  4537 (2002).

\bibitem{Schack2001a}
R. Schack, T.~A. Brun, and C.~M. Caves, 
``Quantum Bayes rule'',
{\it Phys.\ Rev.\ A\/} {\bf 64},  art.~no.~014305 (2001).

\bibitem{Brun2001a}
T.~A. Brun, C.~M. Caves, and R. Schack, 
``Entanglement purification of unknown quantum states'',
{\it Phys.\ Rev.\ A\/} {\bf 63},  art.~no.~042309 (2001).

\bibitem{Hardy-0101}
L. Hardy, ``Quantum theory from five reasonable axioms'', 
e-print quant-ph/0101012 v4 (25 Sep 2001).

\bibitem{Hardy-0111}
L. Hardy, ``Why quantum theory?'',
in {\em Proceedings of the NATO Advanced Research Workshop on
  Modality, Probability, and Bell's theorem}, edited by J. Butterfield and T.
  Placek (IOS Press, Amsterdam, 2002), e-print quant-ph/0111068.

\bibitem{VonMises1957}
R. von Mises, {\em Probability, Statistics and Truth\/} 
(Dover, New York, 1957).

\bibitem{vanFraassen1977}
B.~C. van Fraassen, 
``Relative frequencies'', {\it Synthese\/} {\bf 34},  133  (1977).

\bibitem{Jaynes2003}
E.~T. Jaynes, {\em Probability Theory\/} 
(Cambridge University Press, Cambridge, 2003).

\bibitem{Hardy-private}
L. Hardy, private communication. See also remarks
at the end of section 6.1 of \cite{Hardy-0101}.

\bibitem{Fuchs2002a}
C.~A. Fuchs, ``Quantum states: What the hell are they?'', unpublished (2002), 
p.~159--166, available at http://cm.bell-labs.com/who/cafuchs.

\bibitem{Fuchs-0205}
C.~A. Fuchs, 
``Quantum mechanics as quantum information (and only a little more)'',
e-print quant-ph/0205039.

\end{thebibliography}

\end{document}